\documentclass[aps,pre,twocolumn,showpacs,floatfix,superscriptaddress]{revtex4}
 \usepackage{graphicx}
\usepackage{epsfig}

\newcommand{\bx}{{\bf x}}
\newcommand{\bv}{{\bf v}}

\begin{document}

\title{Optimizing Replica Exchange Moves For Molecular Dynamics}  

\author{Walter Nadler}
\email{wnadler@mtu.edu, w.nadler@fz-juelich.de}
\affiliation{John-von-Neumann Institute for Computing, 
             Forschungszentrum J\"ulich, D-52425 J\"ulich, Germany}

\author{Ulrich H. E. Hansmann}
\email{hansmann@mtu.edu, u.hansmann@fz-juelich.de}
\affiliation{Department of Physics, Michigan Technological University, Houghton, Michigan, USA}
\affiliation{John-von-Neumann Institute for Computing, 
             Forschungszentrum J\"ulich, D-52425 J\"ulich, Germany}

\date{\today}

\begin{abstract}
In this short note we sketch the statistical physics framework of the replica exchange technique when applied to molecular dynamics simulations.
In particular, we draw attention to generalized move sets that allow a variety of optimizations as well as new applications of the method.
\end{abstract}

\maketitle

\begin{figure}[t]
\epsfig{file=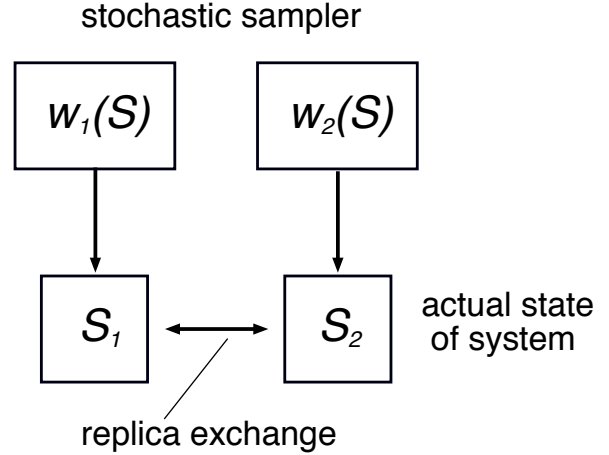,width=6cm, angle=90}
\caption{ Sketch of the basic situation in replica exchange simulations:
simulations run independently on stochastic samplers with weight function $w_i(S)$; at certain times exchange of the current conformations is attempted with probability $p_M$, Eq.~(\protect\ref{eq:MetropolisGeneral}).
\label{fig1}
}
\end{figure}

Effective simulation of proteins~\cite{SK2001}, glasses~\cite{Binder}, and similar complex systems~\cite{Complex} is 
hampered by slow relaxation due to  barriers and bottlenecks.
Replica exchange \cite{Geyer1995,HN,H97f} - also known as parallel tempering  - is one of the main approaches to overcome these problems.
Originally devised for stochastic simulations, it is  used nowadays also in combination with molecular dynamics (MD) simulations, i. e. simulations that have a strong deterministic character. We have found that there exists some confusion on the correct application of the replica exchange approach to MD. In this short note we will sketch the theoretical basis for applying replica exchange to MD simulations, introduce generalized move sets necessary for optimizing exchange rates, and  point to possible extensions of the concept.

In replica exchange, a set of stochastic simulations are performed 
in parallel with distinct weight functions $w_i(S)$, see Fig.~1.
At certain times an exchange of 
 current conformations $S_i$ of replicas at neighboring samplers \cite{footnote1} is attempted, e.g. for a single pair of simulations
\begin{equation}
\left\{S_1,S_2\right\} \to \left\{S_1',S_2'\right\}=\left\{S_2,S_1\right\} 
\quad .
\label{eq:MS0}
\end{equation}
Such an exchange move does not change the statistics of the full distribution function
\begin{equation}
w_{tot}(S_1,S_2)=w_1(S_1)w_2(S_2)
\quad ,
\end{equation}
if it  is accepted or rejected according to a generalized Metropolis rule~\cite{Metropolis},
\begin{eqnarray}
p_M\left(\left\{S_1,S_2\right\} \to \left\{S_1',S_2'\right\}\right) &=&
{\rm min} \left( 1,{w_{tot}(S_1',S_2')\over w_{tot}(S_1,S_2)}\right) 
\quad .
\nonumber\\
\label{eq:MetropolisGeneral}
\end{eqnarray}

In physical and chemical applications 
one usually focuses on the canonical ensemble. Using replica exchange an individual replica  performs a random walk in temperature, allowing it to enter and escape local potential minima. As a consequence, the state space is explored more thoroughly, especially at low temperatures. The weight functions employed in such situations are the canonical distributions, $w(S)\propto\exp\left[-\beta E(S)\right]$, with $E(S)$ the energy of the system and $\beta = 1/k_BT$ the inverse temperature. For the exchange move Eq.~(\ref{eq:MS0}), the  Metropolis criterion assumes the simple form
\begin{equation}
p_M\left(\left\{S_1,S_2\right\} \to \left\{S_2,S_1\right\}\right) =
{\rm min} \left( 1,\exp\left[\Delta\beta\Delta E\right]\right)
\label{eq:MetropolisSimple}
\end{equation}
with $\Delta\beta=\beta_2-\beta_1$ and $\Delta E=E(S_2)-E(S_1)$.

The exchange move of Eq.~(\ref{eq:MS0}) is not unique. 
More general moves can be derived involving both 
 exchange $and$ some modification of the exchanged states
\begin{equation}
\left\{S_1,S_2\right\} \to \left\{S_1'(S2),S_2'(S1)\right\} \quad .
\label{eq:MS1}
\end{equation}
Such moves are allowed as long as they preserve  detailed balance and - in combination with the independent simulations - do not violate global ergodicity.  
The exponent of the acceptance probability 
for such generalized replica exchange moves 
is no longer given by the simple form in Eq.~(\ref{eq:MetropolisSimple}) but by

\begin{samepage}
\begin{eqnarray}
\ln\left( W \right) &\equiv& \ln\left[{w_{tot}(S_1',S_2')\over w_{tot}(S_1,S_2)}\right]
\nonumber\\
&=& \beta_1 \left[ (E(S_1)-E(S_1') \right] + 
\beta_2 \left[ E(S_2)-E(S_2') \right] \quad .
\nonumber\\
\label{eq:lnW}
\end{eqnarray}
\end{samepage}

Molecular dynamics, i.e. determining the time evolution of a classical many-body system by numerically solving the equations of motion, is an intrinsically deterministic approach.
At first glance, this property seems to preclude its incorporation into the replica exchange scheme sketched above, since no stochastic sampling appears to be involved. However, already early on MD simulations were viewed also as a stochastic sampling procedure by considering the velocity field as a heat bath \cite{Allen,Frenkel}: A continuous appropriate rescaling of the velocities leads to a correct canonical sampling of all properties depending only on the coordinates of the particles~\cite{Nose,footnote3}. In this way the system state is defined by the coordinates only, $S=\bx$, and the velocity field in a way acts as stochastic sampler. The above replica exchange scheme can then be applied straightforwardly. Care has to be taken, however, to properly treat the velocity field  as stochastic sampler also in this situation, e.g. like in Ref.~\onlinecite{H97f} by randomly reinitializing the velocities upon exchange according to
\begin{equation}
\left< E_{kin} \right> = \frac{1}{2} N k_B T \quad ,
\label{eq:kBT}
\end{equation}
with $N$ being the number of degrees of freedom.

By considering the velocity field as a heat bath one gives up the idea of a trajectory. For this and other~\cite{footnote3}
reasons, it is more common in canonical MD simulations to control the system
by an additional thermostat algorithm \cite{Allen,Frenkel,Nose}.
In such an implementation of canonical MD the system state is given by coordinates $and$ velocities together, $S=(\bx,\bv)$, the thermostat acting as the external stochastic sampler. Consequently, in replica exchange moves  
the full state of the system, i.e  $S=(\bx,\bv)$, has to be considered. The contributions to the energy can be separated as
\begin{equation}
E(S)=E_{pot}(\bx)+E_{kin}(\bv) \quad .
\label{eq:E}
\end{equation}
As a consequence, the  acceptance probability of  Eq.~(\ref{eq:lnW})  takes now on the form
\begin{equation}
\ln(W)=\Delta\beta\Delta E_{pot} + \Delta K 
\label{eq:lnWcan}
\end{equation}
with
\begin{equation}
\Delta K =
(\beta_2 - \beta_1) \left[E_{kin} (\bv_2) - E_{kin}(\bv_1)\right] ~.
\label{eq:DK}
\end{equation}
Since the kinetic energy reflects the simulation temperature, see Eq.~(\ref{eq:kBT}),  in general
$\Delta K$ will be negative. Hence,
a naive application of  Eq.~(\ref{eq:MS0}) is hampered by 
 a large detrimental contribution from the kinetic energy difference 
 yielding a very low   acceptance probability, Eq.~(\ref{eq:MetropolisSimple}).
 Moreover, accepted moves also lead to velocities that are not characteristic 
 for the new temperatures, pushing the system out of equilibrium.

One way to avoid such problems is to turn to generalized exchange moves
that control the possibly large fluctuations in the velocity fields.
A re-scaling of the velocity fields in the  move
\begin{equation}
\left\{(\bx_1,\bv_1),(\bx_2,\bv_2)\right\} \to 
\left\{(\bx_2,r^{-1}\bv_2),(\bx_1,r\bv_1)\right\} 
\quad ,
\label{eq:MSMD}
\end{equation}
leads now to a contribution of the kinetic energy to the acceptance probability 
of Eq.~(\ref{eq:lnWcan}) that is given by 
\begin{equation}
\Delta K(r) =
(\beta_2 - \beta_1 r^{-2}) (E_{kin} (\bv_2) - r^2 E_{kin}(\bv_1)) ~.
\label{eq:DKr}
\end{equation}
where we have used the property $E_{kin}(r\bv)=r^2E_{kin}(\bv)$. 
In order to optimize the acceptance of replica exchange moves one can 
adjust the scale $r$  in such a way that the kinetic energy contribution to the 
Metropolis term vanishes.  The condition $\Delta K(r) = 0$ has two solutions. 
Following Ref.~\cite{SO97} one can choose
\begin{equation}
 r = r_\beta \equiv \sqrt{ \frac{\beta_1}{\beta_2} }  = \sqrt{\frac{T_2}{T_1}} ~.
 \label{Okamoto}
 \end{equation}
 This choice of $r$ leads to $\beta_2 - \beta_1 r^{-2} = 0$ in Eq.~(\ref{eq:DKr}), 
but obviously this is not the only possibility. 
An alternative is to adjust $r$ in a way that
$  E_{kin} (\bv_2) - r^2 E_{kin}(\bv_1) = 0$.  This can be realized by setting
\begin{equation}
r=r_E\equiv\sqrt{E_{kin}(\bv_2)/E_{kin}(\bv_1)}  = \sqrt{\hat{T}_2/\hat{T}_1} \quad .
\label{eq:rE}
\end{equation}
Here, $\hat{T} = 2 k_B E_{kin}$ is the instantaneous temperature of the velocity field, as opposed to the thermostat temperature $T$. 
Obviously, in the thermodynamic limit $\hat{T} \to T$, and both scalings become identical. 
However, in finite systems $\hat{T}$ fluctuates around $T$ and, in general, one has $r_E \ne r_\beta$. 

As both scalings preserve detailed balance~\cite{footnote2}, 
they  lead to equally valid but different replica exchange moves.
Both scalings are optimal in the sense that they render acceptance rates independent of fluctuations in the velocity field.
Current implementations usually employ only a single variant. 
Choosing randomly among both moves can increase mixing in replica exchange, albeit without further increasing acceptance rates.

Since $\Delta K(r) \to -\infty$ for $r\to0$ as well as for $r\to\infty$, 
and $\Delta K(r_\beta) =0 =\Delta K(r_E) $, 
there exists a regime of $r$ where $\Delta K(r)>0$. 
It is therefore tempting to maximize $\Delta K(r)$ 
as this will increase  the acceptance probability of an exchange move. 
From the condition
\begin{equation}
\frac{d}{dr} \Delta K(r)= 
\beta_2 E_{kin}(\bv_1) - \frac{\beta_1 E_{kin}(\bv_2)}{r^4} \equiv 0~,
~,
\label{eq:DK0}
\end{equation}
one finds
\begin{equation}
r_{opt} = \sqrt[4]{\frac{\beta_1 E_{kin}(\bv_2)}{\beta_2 E_{kin}(\bv_1)}}
             = \sqrt{r_{\beta}r_E}~,
 \label{eq:ropt}
   \end{equation}
 which leads to a positive maximal contribution of the kinetic energy given by
\begin{equation}
\Delta K(r_{opt}) =
\left(\sqrt{\beta_1 E_{kin}(\bv_1)}-\sqrt{\beta_2 E_{kin}(\bv_2)}\right)^2~.
\label{eq:Kopt}
\end{equation}
Hence, with such a re-scaling of velocities  the fluctuations of the instantaneous temperatures of the velocity fields can be utilized to $increase$ the acceptance probability of exchange moves, allowing for larger jumps in potential energy. Note, however,
that the re-scaling according to Eq.~(\ref{eq:ropt}) preserves detailed balance only in the thermodynamic limit, i.e. where $\hat{T} \to T$ holds and $\Delta K(r_{opt})\equiv0$ anyhow.
For finite systems and during 
equilibration, i.e. where $\hat{T}$ deviates from $T$ and $\Delta K(r_{opt})>0$, detailed balance is violated, albeit to a lesser degree the larger the system is.
For this reason, the scaling of Eq.~(\ref{eq:ropt})
has to be used with care.

\begin{figure}[t]
\epsfig{file=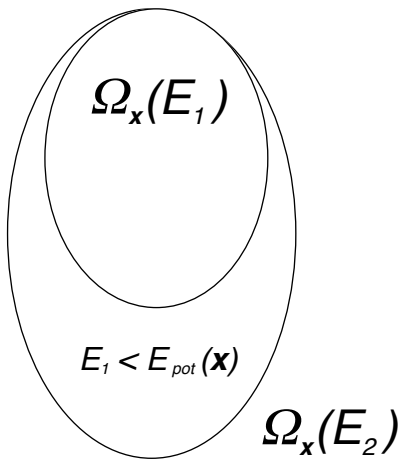,width=6cm,angle=90}
\caption{ Sketch of the coordinate part of the phase space for constant energy replica exchange simulations. Exchange moves are possible in the overlap region.
\label{fig2}
}
\end{figure}

The above theoretical framework allows us to introduce also an interesting new variant of replica exchange MD. Here the stochastic sampler is just the regular MD simulation without a heat bath. Provided the dynamics is ergodic, a microcanonical MD simulation can be viewed as a constant sampling on the energy surface, $E=E_{kin}+E_{pot}=const$. The corresponding weight function is $w(S)=1/|\Omega(E)|$, where $\Omega(E)$ denotes the phase space of the hypersurface of constant energy at $E$. It is well-known that microcanonical MD exhibits slow equilibration, and independently sampled trajectories should be combined to ensure better statistics. However, this approach can be extended readily into a replica exchange scheme for speeding up equilibration on several energy surfaces together. We assume $E_1<E_2$ in the following. The move set is a generalization of Eq.~(\ref{eq:MSMD})
\begin{equation}
\left\{(\bx_1,\bv_1),(\bx_2,\bv_2)\right\} \to 
\left\{(\bx_2,r_2\bv_2),(\bx_1,r_1\bv_1)\right\} 
\quad ,
\label{eq:MSMDE}
\end{equation}
involving two different rescaling factors $r_1$ and $r_2$
\begin{equation}
r_{1,2}=
\sqrt{{E_{2,1}-E_{pot}(\bx_{1,2}) \over E_{1,2} - E_{pot}(\bx_{1,2}) }}\quad .
\label{eq:r12}
\end{equation}
Such moves are possible for $E_{pot}(\bx_2)<E_1$, see Fig.~2; $E_{pot}(\bx_1)<E_2$ automatically holds. We note that this restriction does not violate detailed balance. Furthermore, the combination with the regular MD simulation ensures ergodicity. 

A fascinating aspect of this scheme is that the acceptance probability for an allowed move is always one, since both weight functions are constant. No other scenario allows for such a high acceptance rate. Using reweighting techniques \cite{reweighting}, canonical properties can be obtained from simulations at several different energy values. Moreover, constant energy surface simulations may be of interest in their own right \cite{Allen,Frenkel}, e.g. for comparison with recent molecular beam experiments \cite{Jarrold}.

\begin{figure}[t]
\epsfig{file=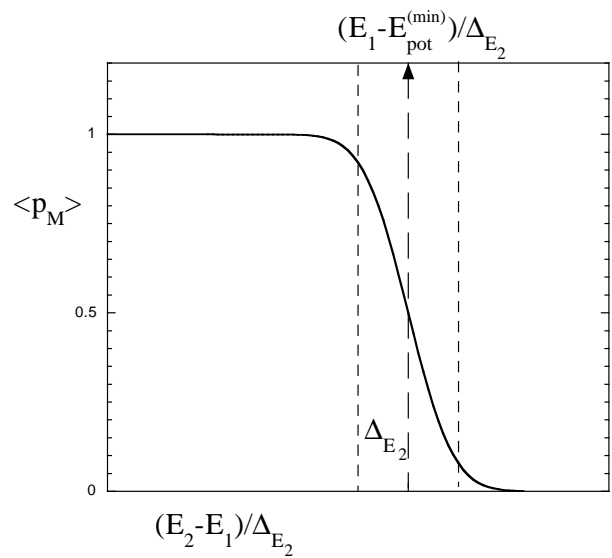,width=8cm}
\caption{ 
Dependence of the average acceptance probability $\left<p_M\right>$ on the distance of the energy shells, see Eq.~(\ref{eq:pMapprox}), in microcanonical replica exchange simulations.
\label{fig3}
}
\end{figure}

Practical acceptance probabilities will be somewhat smaller than one, 
$\left<p_M\right> <1$, due to the region of forbidden moves. Using the distribution of the potential energy on energy shell $E_i$, $P_{E_i}(E_{pot})$,
the average acceptance probability is given by 
\begin{equation}
\left<p_M\right> =
\int_{E_{pot}^{(min)}}^{E_1} dE_{pot} P_{E_2}\left(E_{pot}\right) \quad .
\label{eq:pMgeneral}
\end{equation}
For classical trajectories of total energy $E$ equipartition of average kinetic and potential energy usually holds, 
\begin{equation}
\left<E_{kin}\right>_E=\left<E_{pot}-E_{pot}^{(min)}\right>_E=
\left(E-E_{pot}^{(min)}\right)/2 \quad ,
\label{eq:equipartition}
\end{equation}
with $E_{pot}^{(min)}$ the energy of the lowest energy configuration. 
Assuming a Gaussian distribution for the potential energy, 
\begin{equation}
P_{E_i}(E_{pot})\propto 
\exp\left[-\left(E_{pot}-\left<E_{pot}\right>_{E_i}\right)/
\Delta_{E_i}^2\right] \quad , 
\label{eq:PEi}
\end{equation}
we obtain the dependence of the average acceptance probability $\left<p_M\right>$ on the energy difference $E_2-E_1$ 
\begin{equation}
\left<p_M\right> \approx \frac{1}{2}
{\rm erfc} \left(\frac{E_2-E_1}{\Delta_{E_2}}-
\frac{E_1-E_{pot}^{(min)}}{\Delta_{E_2}}\right)\quad 
\label{eq:pMapprox}
\end{equation}
shown in Fig.~3.
In particular, $\left<p_M\right>$  will be at least one half if the average potential energy at $E_2$ is smaller than $E_1$, $\left<E_{pot}\right>_{E_2}<E_1$. This criterion is equivalent to the energy difference being at most equal to twice the average kinetic energy at $E_1$, $E_2-E_1<E_1-E_{pot}^{(min)}=2\left<E_{kin}\right>_{E_1}$.
Figure 3 shows that $\left<p_M\right>$ will rapidly approach one upon decreasing energy difference.

This last example demonstrates the wide applicability of generalized replica exchange move sets. It also demonstrates the striking advantage of replica exchange over earlier approaches like simulated tempering \cite{ST}. In the latter, additional parameters reflecting free energy differences are of utmost importance. Their determination is tedious and approximations \cite{Pande} are useful only in certain limiting cases.
In replica exchange such normalization constants simply drop out due to the form of the acceptance probability, Eq.~(\ref{eq:MetropolisGeneral}).
Constant energy surface simulations as sketched above could be approached in simulated tempering only with a solid knowledge of the phase space ratios, $|\Omega(E_2)|/|\Omega(E_1)|$.

In summary, we have sketched the statistical physics framework of applying the replica exchange technique to MD simulations. Generalized move sets, in particular appropriate rescaling of the velocity fields, allow optimization of the acceptance probability as well as new approaches.
Together with an optimization of the temperature spacing to increase replica 
flow \cite{Trebst2006a,NH2007} 
optimized acceptance probabilities will lead to shorter simulation times in canonical replica exchange MD simulations.
Microcanonical replica exchange MD simulations are intrinsically optimized and will provide new insights.

\begin{acknowledgments}
It is a pleasure to thank S. Hoefinger and P. Grassberger for discussions.
This research was supported by NSF-grant No. CHE-0313618.
\end{acknowledgments}



\begin{thebibliography}{99}
\bibitem{footnote2}      
Any state-independent scale $r$ preserves detailed balance (DB). In addition, there exist two classes of state-dependent scalings $r=r\left(\bv_1,\bv_2\right)$ that also preserve DB: One is $r=r_E$, Eq.~(\ref{eq:rE}), and the other is any function of a product of the kinetic energies, $r=f\left[E_{kin}(\bv_1)E_{kin}(\bv_2)\right]$. Of all scalings that are compatible with DB, only $r=r_\beta$ and $r=r_E$ are distinguished in that they lead to $\Delta K(r)\equiv 0$.
\end{thebibliography}
\end{document}